\documentclass[11pt,preprint]{aastex}

\usepackage{hyperref}
\hypersetup {
colorlinks=true,
urlcolor=blue
}

\lefthead{van Dokkum et al.}
\righthead{}
\slugcomment{}
\begin{document}

\title{3D-HST Data Release v3.0}


\author{
Pieter van Dokkum\altaffilmark{1},
Gabriel Brammer\altaffilmark{2},
Ivelina Momcheva\altaffilmark{1},
Ros Skelton\altaffilmark{1},
Katherine E.\ Whitaker\altaffilmark{3},
for the 3D-HST team}

\altaffiltext{1}
{Department of Astronomy, Yale University, New Haven, CT 06511, USA}
\altaffiltext{2}
{European Southern Observatory, Alonson de C\'ordova 3107, Casilla
19001, Vitacura, Santiago, Chile}
\altaffiltext{3}
{Astrophysics Science Division, Goddard Space Center, Greenbelt,
MD 20771, USA}

\begin{abstract}

3D-HST is a 248-orbit
Treasury program to provide WFC3 and ACS grism spectroscopy over four
extra-galactic fields (AEGIS, COSMOS, GOODS-South, and UDS), augmented
with previously obtained data in GOODS-North.
We present a new data release of the 3D-HST survey, version v3.0. This
release follows the initial v0.5 release that accompanied the
survey description paper (Brammer et al. 2012). The new v3.0 release includes
the deepest near-IR HST grism spectra currently in existence, extracted
from the 8-17 orbit depth 
observations in the Hubble Ultra Deep Field. Contamination-corrected
2D and 1D spectra, as well as derived redshifts, are made available
for $>250$ objects in this $2'\times 2'$ field. 
The spectra are of extraordinary quality,
and show emission features in many galaxies
as faint as F140W\,$=26-27$, absorption features in quiescent
galaxies at $z\sim 2$, and several active galactic nuclei.
In addition to these extremely deep grism data
we provide reduced WFC3 F125W, F140W, and
F160W image mosaics of all five 3D-HST/CANDELS fields.

\end{abstract}

\section{Overview}

3D-HST is an HST Treasury program executed in Cycles 18 and 19. The survey uses
the ACS and  WFC3 grisms, providing slitless
spectroscopy for all objects in the fields of these instruments. The primary
and most innovative aspect of the survey is 2-orbit depth WFC3/G141 spectroscopy,
covering the wavelength range $1.1\,\mu{}$m -- 1.65\,$\mu$m. 
Owing to the low sky background from space and the excellent sensitivity of
WFC3 we reach a point-source
continuum depth of $H_{140}\sim 23$ at $5\sigma$ in 2 orbits.
In addition to the G141 grism data the 3D-HST data comprise
direct WFC3 images in the $H_{140}$ filter, parallel ACS G800L grism
exposures, and direct images in the ACS $I_{814}$ filter.

The survey fields are four of the fields of the CANDELS Multi-Cycle Treasury
program: GOODS-South, UDS, AEGIS, and COSMOS. The fifth CANDELS field,
GOODS-North, was observed by program GO-11600 (PI: Weiner) in Cycle 17 to
the same 2-orbit G141 depth as 3D-HST (but without parallel ACS grism data).
We have included these GOODS-North data in our analysis.
CANDELS is primarily an imaging survey, providing $J_{125}$ and $H_{160}$
data at two different depths over the five fields. The 3D-HST
grism data cover approximately 80\,\% of the area of the CANDELS fields,
and form a spectroscopic complement to the CANDELS imaging.

The reduction, analysis, and interpretation of the slitless spectroscopy is not
straightforward. The spectra of neighboring objects can overlap, which means a
full 2D model of all spectra needs to be constructed and subtracted prior to
analyzing the object of interest. Furthermore, the ``PSF'' of the grism
spectra is effectively the (wavelength-dependent) morphology of the galaxy,
complicating the fitting procedure. STScI provides
software\footnote{http://axe-info.stsci.edu} to analyze grism spectra but
this is not optimized for faint-object spectroscopy; we therefore developed
fully independent, custom packages that enable optimal modeling and fitting
of interlaced 2D spectra. We find that these tools
work very well, even on the deepest (17-orbit) grism data that
are currently available (see below).

A further complication is that the grism spectra are often
difficult to interpret without information from other wavelengths, as
correct identification of faint emission lines usually requires some
prior information on the likely redshifts of the objects. 
Data at other wavelengths are also crucial for measuring stellar masses,
rest-frame colors, star formation rates, and other parameters.
With these goals in mind we have created photometric catalogs in the 3D-HST
fields, as a first and necessary step to interpret the grism spectra.

\section{Data release v3.0: extremely deep spectroscopy in the UDF
and WFC3 mosaics}

\subsection{Spectroscopy}

We are providing the
G141 grism spectra in the only region of the survey which was observed
multiple
times: the Hubble Ultra Deep Field. We have used this field to test our
fitting procedures and to extract and analyze the deepest near-IR
spectra of faint galaxies currently available. The G141 spectra are based
on a combination of 8 orbits of 3D-HST data and 9 orbits of CANDELS
supernova follow-up data. An analysis of the 17-orbit spectrum of a
candidate $z\sim 12$ galaxy in this field was presented in Brammer et al.\
(2013).
We provide $>250$ spectra and redshifts
over the $2' \times 2'$
UDF. This density of $>60$ spectra per arcmin$^2$ far exceeds what
has been achieved in any other survey or can reasonably be done
from the ground. 

\subsection{WFC3 mosaics}

As part of our commitment to release all ancillary data used in the
3D-HST program
we are providing our reduced WFC3 mosaics of all five 3D-HST/CANDELS fields
as part of data release v3.0.
The mosaics include the deep and wide $J_{125}H_{160}$ data obtained
as part of the CANDELS program,
the 3D-HST $H_{140}$ imaging, and
the Early Release Science observations (and UDF flanking fields) in
GOODS-South.
The CANDELS project has also released several of the imaging mosaics.
Our reduction uses the same tangent
point and pixel scale as CANDELS, which means 3D-HST and CANDELS
objects can be trivially matched and compared.
We are preparing to release our photometric catalogs in these five fields,
including derived photometric redshifts and stellar population parameters.
The catalog release (v3.1) will be accompanied by a paper describing the
reduction of the mosaics and the making of the catalogs (Skelton et al.,
in preparation).

\section{Links}
Information on the 3D-HST survey, as well as links to the data release,
publications, pointing layout, and other information
can be obtained from the 3D-HST website:
\begin{itemize}
\item
{\bf \href{http://3dhst.research.yale.edu/}{http://3dhst.research.yale.edu}}
\end{itemize}
The data release page and the accompanying
{\bf \href{http://monoceros.astro.yale.edu/RELEASE_V3.0/Photometry/3dhst_v3.0_readme.pdf}{comprehensive release notes}}
can also be accessed directly:
\begin{itemize}
\item
{\bf
\href{http://3dhst.research.yale.edu/Data.html}{http://3dhst.research.yale.edu/Data.html}}
\end{itemize}
We have created several tools to explore the data release.
For the Ultra Deep Field we provide an HTML table with image thumbnails
and spectra, which can be ordered according to ID, magnitude or redshift.
An interactive image browser displays grism redshifts on UDF objects
and allows users to scroll around the
field using a google maps-like interface. Clicking on an object
brings up the HTML table with the object at the top.
\begin{itemize}
\item
{\bf
\href{http://monoceros.astro.yale.edu/RELEASE_V3.0/Spectra/UDF/Web/UDF_3dhst_redshift_v1.0.html}{UDF HTML catalog with RGB thumbnails, spectra, and fits}}
\item
{\bf
\href{http://monoceros.astro.yale.edu/RELEASE_V3.0/Spectra/UDF/Web/HUDF_iJH.html#03:32:39.73,-27:46:11.3}{UDF interactive image browser}}
\end{itemize}
Image browsers for all five fields can be used for roaming the 3D-HST/CANDELS
fields or to go directly to particular locations by entering coordinates:
\begin{itemize}
\item
{\bf
\href{http://monoceros.astro.yale.edu/RELEASE_V3.0/Photometry/RGB/AEGIS3/AEGIS_iJH.html#14:20:19.49,+53:02:03.8}{AEGIS $I_{814}J_{125}H_{160}$ RGB mosaic}}
\item
{\bf
\href{http://monoceros.astro.yale.edu/RELEASE_V3.0/Photometry/RGB/COSMOS3/COSMOS_iJH.html#10:00:17.97,+02:18:05.5}{COSMOS $I_{814}J_{125}H_{160}$ RGB mosaic}}
\item
{\bf
\href{http://monoceros.astro.yale.edu/RELEASE_V3.0/Photometry/RGB/GS3/GS_iJH.html#03:32:14.96,-27:42:25.9}{GOODS-South $I_{814}J_{125}H_{160}$ RGB mosaic}}
\item
{\bf
\href{http://monoceros.astro.yale.edu/RELEASE_V3.0/Photometry/RGB/GN3/GN_iJH.html#12:37:21.09,+62:12:46.8}{GOODS-North $I_{775}J_{125}H_{160}$ RGB mosaic}}
\item
{\bf
\href{http://monoceros.astro.yale.edu/RELEASE_V3.0/Photometry/RGB/UDS3/UDS_iJH.html#02:17:37.17,-05:13:30.1}{UDS $I_{775}J_{125}H_{160}$ RGB mosaic}}
\end{itemize}

\begin{acknowledgements}
\noindent
{\bf Acknowledgements:} 
The 3D-HST survey, programs HST-GO-12177 and HST-GO-12328,
is based on observations made
with the NASA/ESA Hubble Space Telescope, obtained at the Space
Telescope Science Institute, which is operated by the Association of
Universities for Research in Astronomy, Inc., under NASA contract NAS
5-26555. Financial support for this program is gratefully
acknowledged. Besides data from the 3D-HST survey itself we used
observations from programs 11600 (PI: Weiner), CANDELS (12060-12064 and
12440-12445; PIs: Faber,
Ferguson), the WFC3 Early Release Science Program (11359; PI: O'Connell),
GOODS (9425 and 9573; PI:
Giavalisco), HUDF-09 (11563; PI: Illingworth), and
HUDF-12 (12498; PI: Ellis) as delivered by the HST Archive.
We thank our colleagues for
providing these extraordinary data to the community.

\end{acknowledgements}


\begin{references}


\reference{} {Brammer}, G.~B., {van Dokkum}, P.~G., {Franx}, M., {Fumagalli}, M., {Patel},  S., {Rix}, H.-W., {Skelton}, R.~E., {Kriek}, M., {et al.} 2012, \apjs, 200, 13

\reference{} {Brammer}, G.~B., {van Dokkum}, P.~G., {Illingworth}, G.~D.,
{Bouwens}, R.~J., {Labb\'e}, I., {Franx}, M., {Momcheva}, I., \& {Oesch},
P.~A. 2013, \apj, 765, L2





\end{references}
\end{document}